\documentclass[runningheads]{llncs}
\usepackage{graphicx}
\usepackage{amsmath}
\usepackage{amssymb}
\usepackage{tabularx,longtable,multirow}
\usepackage{color}
\usepackage{tabularx}
\newcolumntype{C}[1]{>{\centering\arraybackslash}p{#1}}
\begin{document}
\title{D-Net: Siamese based Network with Mutual Attention for Volume Alignment}

%
%
\author{Jian-Qing Zheng\inst{1} \and Ngee Han Lim\inst{1}\and Bart{\l}omiej W. Papie{\.z}\inst{2}}

%
\authorrunning{Zheng, et al.}
%
\institute{Kennedy Institute of Rheumatology, Nuffield Department of Rheumatology, Orthapaedics and Musculoskeletal Sciences, University of Oxford, UK\\ \email{\{jianqing.zheng, han.lim\}@kennedy.ox.ac.uk} \and 
Big Data Institute, Li Ka Shing Centre for Health Information and Discovery University of Oxford, Oxford, UK\\ \email{bartlomiej.papiez@bdi.ox.ac.uk}}
\maketitle              
\begin{abstract}
Alignment of contrast and non contrast-enhanced imaging is essential for quantification of changes in several biomedical applications. In particular, the extraction of cartilage shape from contrast-enhanced Computed Tomography (CT) of tibiae requires accurate alignment of the bone, currently performed manually. Existing deep learning-based methods for alignment require a common template or are limited in rotation range. Therefore, we present a novel network, D-net, to estimate arbitrary rotation and translation between 3D CT scans that additionally does not require a prior template. D-net is an extension to the branched Siamese encoder-decoder structure connected by new mutual non-local links, which efficiently capture long-range connections of similar features between two branches. The 3D supervised network is trained and validated using preclinical CT scans of mouse tibiae with and without contrast enhancement in cartilage. 
The presented results show a significant improvement in the estimation of CT alignment, outperforming the current comparable methods.

 

\end{abstract}
\section{Introduction}
It is currently impossible to accurately quantify the damage to cartilage during the progression of disease in small animal models of osteoarthritis. Visualisation of cartilage in Computed Tomography (CT) requires a contrast agent. The preclinical development of such a contrast agent \cite{ngee2019Radiopaque} has highlighted the problem of accurate cartilage shape extraction from contrasted images; the partial volume effect and adjacency to bone necessitates the use of pre-and post-contrasted CT images. In preclinical scanners, the animal can be placed in various unsystematic (i.e. arbitrary) positions during the acquisition. Tibial cartilage shape may be extracted from the contrast enhanced image by subtracting the non-contrasted scan but this requires accurate alignment of the tibial bone.
However, current semi-accurate manual alignment using \textit{ImageJ} requires over 1 hour and is prone to error, calling for an automated and accurate method to estimate rigid transformation between 3D volumes acquired in the preclinical setup. 


The standardised protocols for image acquisition in clinical scanners means that the range of rotation and translation required to register scans are small and the bigger challenge is to perform deformable registration, especially in between image modalities \cite{schnabel2016advances}. In pre-clinical studies, protocols are usually study and machine specific. The limbs of mice are particularly challenging as they may be extended or tucked, dependent on posture (prone, supine and on the side). Post-mortem \textit{ex vivo} tissue may also be scanned in fixative solution, which increases the variability of orientations and positions of scans. Our initial dataset is comprised of such \textit{ex vivo} tissue which will later be used to validate the \textit{in vivo} scans. Thus estimation of large-range rigid transformations is required. 

Classic approaches to preclinical image alignment \cite{baiker2011automated} used state-of-the-art, iterative image registration with a similarity measure capturing intensity changes caused by contrast and an appropriate transformation model. However, such approaches are easily trapped in local minimum especially when large translation or rotation is present. More recently, deep learning approaches \cite{simonovsky2016deep},\cite{liao2017artificial},\cite{ma2017multimodal} were employed to improve the performance of iterative image registration algorithms, however, the slow performance and dependency on initialization motivates one-step transformation estimation via regression \cite{haskins2019learning,haskins2020deep}. For example, two-branch Siamese Encoder (SE) used to learn similarity measure between two images, was applied to 2D brain images alignment \cite{sloan2018learning}. 
A convolution neural network called AIRNet \cite{chee2018airnet} was used for affine registration of 3D brain Magnetic Resonance Imaging (MRI) with dense convolution layers as SE. 
The SE structure was also used within the framework of deformable image registration in \cite{de2019deep,wang2019fire} to estimate an initial, affine transformation between two volumes. 
Alternatively, affine transformation can be estimated using the Global-net \cite{hu2018label} with the input images being concatenated and fed into an one-branch encoder. 
Despite the success of the previous approaches, the capture range of rotation is heavily limited between $\pm$15$^\circ$ \cite{sloan2018learning} and $\pm$45.84$^\circ$ (0.8 rad) \cite{chee2018airnet}, yielding unsatisfactory results in preclinical imaging acquisition setup (shown in Sec.~\ref{sec:result}). 
The 3D pose estimation of arbitrary oriented subject was presented in \cite{salehi2018real} but it requires a prior standard template, which is not available for preclinical cartilage imaging. 

In this paper, a new architecture, D-net, is proposed for estimation of arbitrary rigid transformation based on a Siamese Encoder Decoder (SED) with novel Mutual Non-local Links (MNL) between two Siamese branches, as described in Sec.~\ref{sec:net} and Sec.~\ref{sec:mnl}. Data collection and experiment design are described in Sec.~\ref{sec:data} and Sec.~\ref{sec:strategy} respectively. Experimental results are shown in Sec.~\ref{sec:result}, discussed and concluded in Sec.~\ref{sec:discussion_conclusion}.

The contributions of this work are as follows.
We propose a new network with SED used for first time for rigid registration and we present a concept of MNL  showing significantly improved performance on 3D volume alignment. 
Our network achieves consistent accuracy for wide range of volume orientations apparent in challenging preclinical data set while it does not require prior atlas or template. 


\section{Methodology}
\label{sec:method}
The objective of 3D image registration is to estimate the transformation $f: \mathbb{R}^{\textbf{\textit{s}}}\to\mathbb{R}^{\textbf{\textit{s}}},\textbf{\textit{X}}^{\rm f}\mapsto\textbf{\textit{X}}^{\rm m}$ between a fixed volume
$\textbf{\textit{X}}^{\rm f}\in\mathbb{R}^{\textbf{\textit{s}}}$ and a moving volume $\textbf{\textit{X}}^{\rm m}\in\mathbb{R}^{\textbf{\textit{s}}}$, where $\textbf{\textit{s}}={d\times h \times w}$, and $d,h,w$ are the thickness, height, and width. For 3D rigid registration, the transformation $f_\theta:=[\textit{\textbf{R}},\textit{\textbf{t}}]\in {SE}(3)$, consists of rotation $\textit{\textbf{R}}\in {SO}(3)$ and translation $\textit{\textbf{t}}\in\mathbb{R}^{3}$, with the parameters $\theta=[\theta^{\rm r},\theta^{\rm t}]\in\mathbb{R}^{12}$ including $\theta^{\rm t}$ and $\theta^{\rm r}$ for translation and rotation. 
The task of the networks in registration is to estimate $\theta$ from the two preprocessed volumes $\tilde{\textbf{\textit{X}}}^{\rm f}$ and $\tilde{\textbf{\textit{X}}}^{\rm m}$ by networks' mapping $g: (\tilde{\textit{\textbf{X}}}^{\rm f},\tilde{\textit{\textbf{X}}}^{\rm m})\mapsto\hat{\theta}$, where $\hat{\theta}\in\mathbb{R}^{12}$ are the parameters estimated by networks.

As $\theta^{\rm r}\in\mathbb{R}^{9}$ is redundant for rotation, the 3D orthogonalization mapping of 6D rotation representation \cite{zhou2019continuity} is used as $O:\mathbb{R}^6\to SO(3),~{\theta}^{\rm r}_{1:6}\mapsto{\textbf{\textit{R}}}$, calculated by:
\begin{equation}
O({\theta}^{\rm r}_{1:6})=[\textit{\textbf{r}}_1~\textit{\textbf{r}}_2~\textit{\textbf{r}}_3]:=
\begin{bmatrix}
N([{\theta}^{\rm r}_{1:3}]^\top)\\
N([{\theta}^{\rm r}_{4\--6}]^\top-(\textit{\textbf{r}}_1^\top\cdot[{\theta}^{\rm r}_{4:6}])\textit{\textbf{r}}_1^\top)\\
{\rm det}([\textit{\textbf{r}}_1~\textit{\textbf{r}}_2~\textit{\textbf{e}}])
\end{bmatrix}^\top
\end{equation}
where $[\theta_{i:j}]\in\mathbb{R}^{j-i}$ denotes a column vector consist of $\theta_{i:j}$, $N(\cdot)$ denotes a Euclidean normalization function, ${\rm det}(\cdot)$ denotes a determinant calculation, $\textbf{\textit{e}}$ is a vector of the 3 canonical basis vectors of the 3D Euclidean space.
This mapping keeps the continuous representation of 3D rotation and is equivalent to Gram-Schmidt process for a rotation in right handed coordinate system but just requires 6 input values.

Thus the rigid transformation can be estimated by: $\hat{f}_{\hat{\theta}}=[O(\hat{\theta}^{\rm r}),\hat{\theta}^{\rm t}]$.

\begin{figure}[th]
\includegraphics[width=\textwidth]{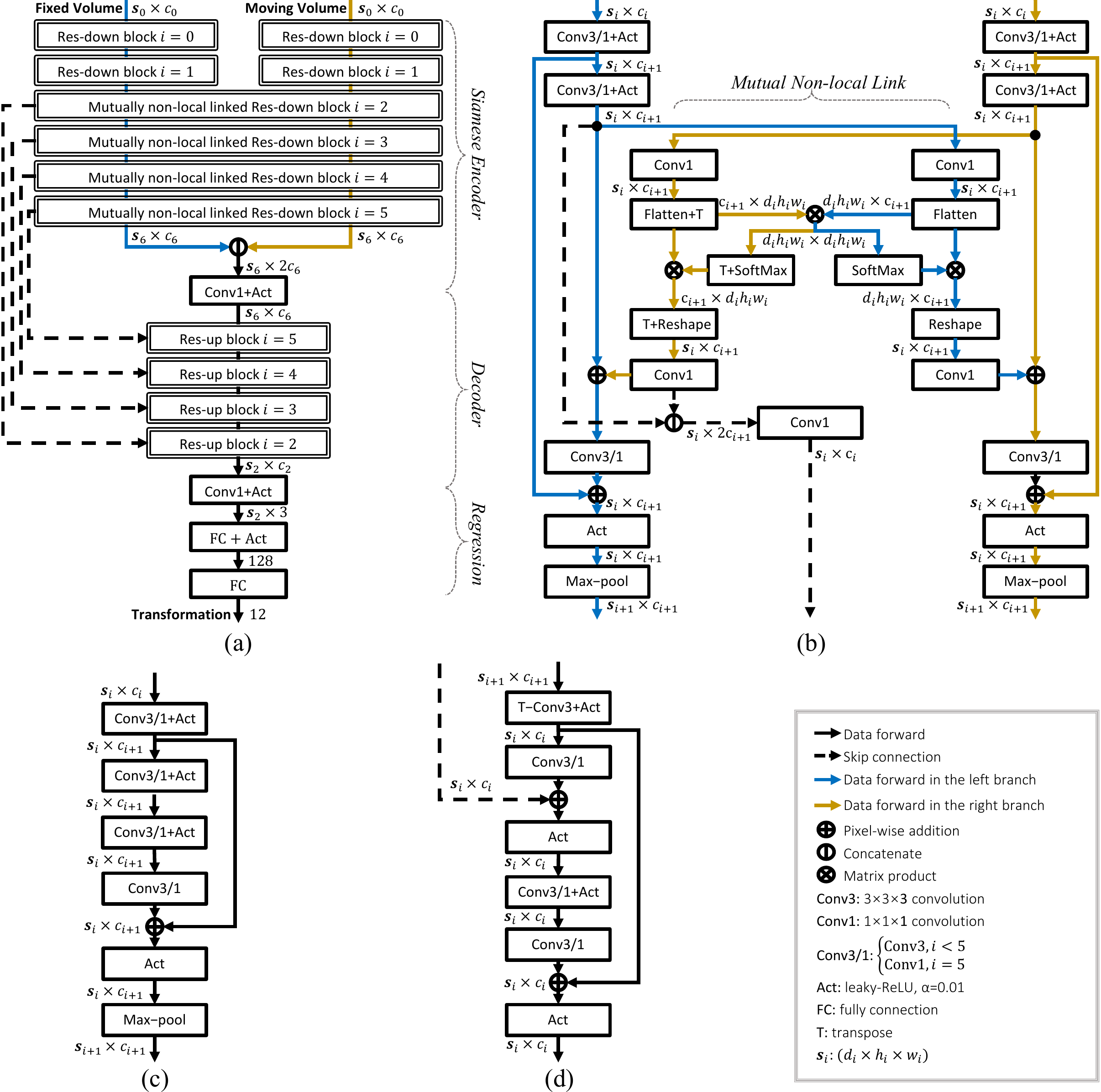}
\caption{The architecture of (a) D-net, (b) the novel Mutually non-local linked Res-down block, (c) Res-down block, and (d) Res-up block.} 
\label{networks}
\end{figure}

\subsection{D-net architecture}
\label{sec:net}

D-net consists of SE part, decoder part, and regression part, and its schematic architecture is shown in Fig.~\ref{networks}(a).
Similar structure of SED was applied to segmentation \cite{kwon2019siamese} and tracking \cite{dunnhofer2020siam}, but with different connection structure between contracting and expansive parts comparing to D-net.
The SE in the D-net includes two branches of six Residual-down-sampling (Res-down) blocks with shared parameters. 
Four pairs of the Res-down blocks are linked by MNL and detailed in the Fig.~\ref{networks}(b). In MNL, two matching matrices, from left branch to right and the inverse, are computed by dot product of each pair of voxels' feature vectors. The matrices from two branches are normalized via softmax to correspond and connect the voxels of the feature maps between two branches. MNL, therefore, captures the long-range connection of similar high and low level features between two branches.  

The details of the Res-down blocks are illustrated in the Fig.~\ref{networks}(c).  
The decoder part of D-net includes four Residual-up-sampling (Res-up) blocks receiving skip connections from the corresponding Mutual Non-local Linked Res-down blocks, shown in the Fig.~\ref{networks}(d). The regression part of D-net includes two fully connected layers with 128 and 12 neurons for 12 transformation parameters.
In Fig.~\ref{networks}, $i$ is the block number, $\textbf{\textit{d}}=(d_0 \cdots d_6)\in\mathbb{Z}_+^{7}$, $\textbf{\textit{h}}=(h_0 \cdots h_6)\in\mathbb{Z}_+^{7}$,$\textbf{\textit{w}}=(w_0 \cdots w_6)\in\mathbb{Z}_+^{7}$ and $\textbf{\textit{c}}=(c_0 \cdots c_6)\in\mathbb{Z}_+^{7}$ denote the sequences of thickness, heights, widths, and channel number of input volume and feature maps for each branch respectively.

\subsection{Mutual Non-local Link}
\label{sec:mnl}

An approach with non-local links was presented in a classic image registration, where non-local motion was estimated using a range of spatial scales, naturally captured by graph representation \cite{papiez2016non}.
Similarly, a concept of unique matching between a pair of voxels by weighting function and mutual saliency was previously shown in \cite{ou2011dramms}.
Here, the mutual attention mechanism is proposed with the deep-learning design of MNL inspired by the Self Non-local Link (SNL) on one branch proposed in \cite{wang2018non}. 
In this section we provide the general definition of MNL:
\begin{equation}
\left\{
\begin{array}{cc}
\textbf{\textit{y}}_k^{\rm m2f}:=\frac{\sum_{\forall j}{\phi({\textbf{\textit{x}}_{k}^{\rm f}},{\textbf{\textit{x}}_{j}^{\rm m}})\psi({\textbf{\textit{x}}_{j}^{\rm m}})}}{\sum_{\forall j}{\phi({\textbf{\textit{x}}_{k}^{\rm f}},{\textbf{\textit{x}}_{j}^{\rm m}})}}\\
\textbf{\textit{y}}_k^{\rm f2m}:=\frac{\sum_{\forall j}{\phi({\textbf{\textit{x}}_{k}^{\rm m}},{\textbf{\textit{x}}_{j}^{\rm f}})\psi({{\textbf{\textit{x}}}_{j}^{\rm f}})}}{\sum_{\forall j}{\phi({{\textbf{\textit{x}}}_{k}^{\rm f}},{{\textbf{\textit{x}}}_{j}^{\rm m}})}}
\end{array}
\right.
\end{equation}
where $j,k$ are the indices of the position in a feature map, $\textbf{\textit{x}}^{\rm f},\textbf{\textit{x}}^{\rm m}$ are the input signals from two branches, $\textbf{\textit{y}}^{\rm m2f},\textbf{\textit{y}}^{\rm f2m}$ are the output signals from this block, $\phi$ is the similarity measurement function, $\psi$ is a unary function.

The instantiated MNL in D-net is based on embedded Gaussian similarity representation with
\begin{equation}
\phi(\textbf{\textit{x}}_1,\textbf{\textit{x}}_2):=e^{{\textbf{\textit{x}}_1}^{\top}{\textbf{\textit{W}}}^{\top}{\textbf{\textit{W}}}\textbf{\textit{x}}_2}    
\end{equation}
and 
\begin{equation}
\psi(\textbf{\textit{x}}):=\textbf{\textit{W}}\textbf{\textit{x}}
\end{equation}
, where $\textbf{\textit{W}}$ is a matrix of trainable weights.

\subsection{Data collection and pre-processing}
\label{sec:data}
A total of 100 ex-vivo micro CT scans of tibiae from 50 mice were acquired using $Perkin~Elmer$, Quantum FX with a resolution of $10\times10\times10\mu{\rm m}^3$/vox, volume size of $512\times512\times512$. Scans varied from $0$ week to $20$ weeks post osteoarthritic surgery. Each tibiae was scanned once pre-contrast and once post-contrast with washout. Due to the handling, the muscles, solution and small broken bone fragments are displaced differently from the tibial bone.

\begin{figure}[ht]
\includegraphics[width=\textwidth]{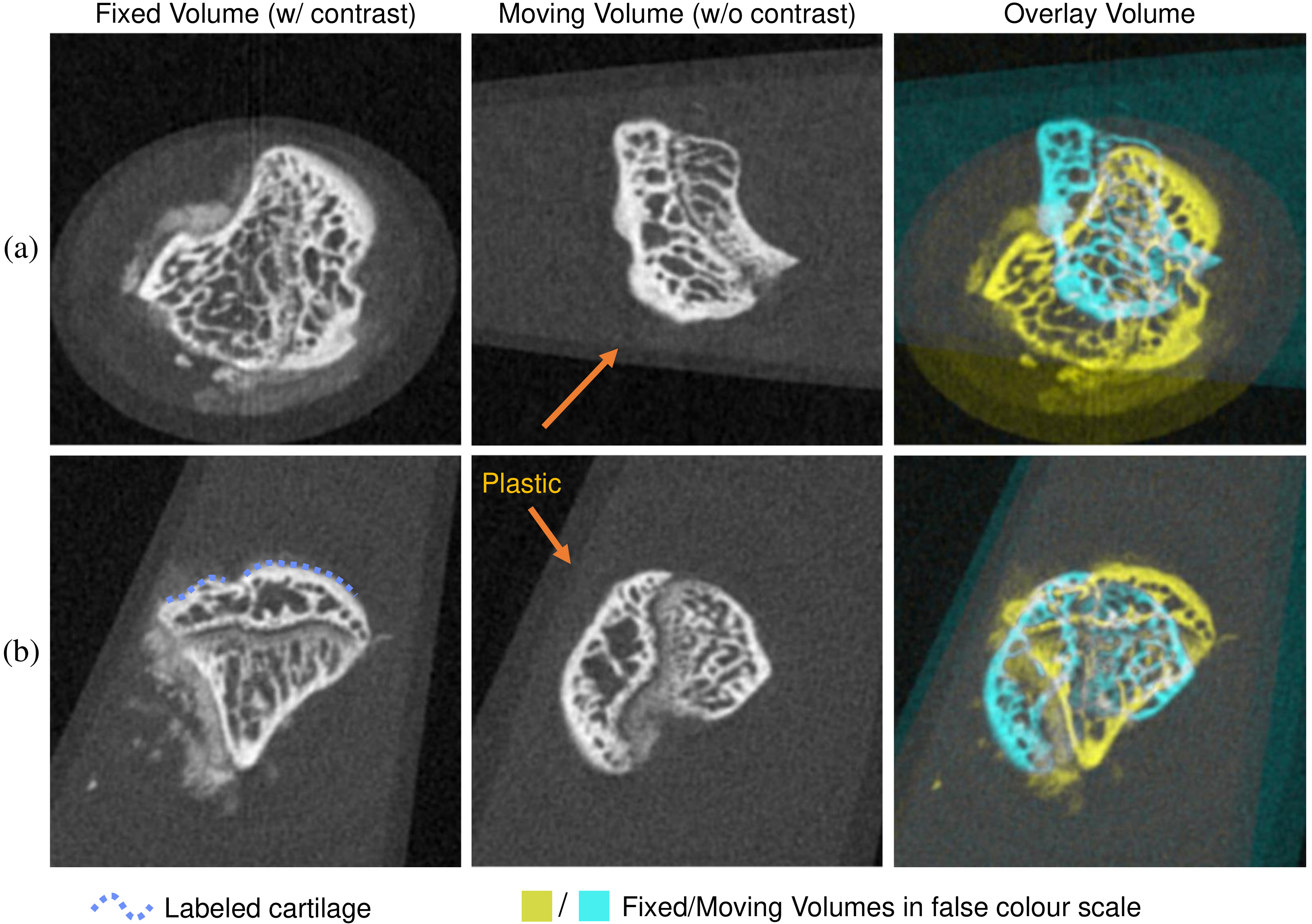}
\caption{The tibial CT slices (at original resolution) of two exemplar subjects show the large ranges of spatial transformation between fixed and moving volumes.} \label{fig:ct}
\end{figure}

To remove the possible influence of muscle and solution on alignment of tibial bones, we preprocessed the collected data by thresholding and normalizing with mapping $T:\mathbb{R}^{\textit{\textbf{s}}}\to\mathbb{R}^{\textit{\textbf{s}}},\textit{\textbf{X}}\mapsto\tilde{\textit{\textbf{X}}}$ by:
\begin{equation}
\tilde{x}_{ijk}=\frac{{\rm ReLU}({x}_{ijk}-{x}_{\rm th})}{x_{\rm max}-{x}_{\rm th}}
\end{equation}
, where $\tilde{x}_{ijk}$ is the entry of $\tilde{\textit{\textbf{X}}}$, ${x}_{\rm max}$ is the maximum intensity in the dataset and ${x}_{\rm th}$ is the threshold value set as 2000 because of the high density of background solution.
Finally, because of data size, the input volumes are sub-sampled with linear interpolation to $80\times80\times80\mu{\rm m}^3$/vox. The CT slices of two exemplar subjects are shown in Fig.~\ref{fig:ct}.

In the training dataset, each CT volume is transformed to synthesize the fixed and the moving volumes with uniformly distributed random translation $\sim\mathcal{U}(-0.64,0.64)$mm in 3D, varying by $\frac{1}{4}$ of the whole volume size, and random rotation with angle $\sim\mathcal{U}(-\pi,\pi)$ around random axes uniformly distributed in 3D sphere surface. To enlarge the training dataset, data augmentation including intensity scale transforming and Gaussian noise is applied, where the intensity scale coefficient is $\sim\mathcal{U}(0.95,1.05)$ and each voxel is added with random variable $\sim\mathcal{N}(0,0.001)$.

\subsection{Training and validation}
\label{sec:strategy}
The loss function in terms of $\theta$ and $\hat{\theta}$ is calculated as: 
\begin{equation}
\mathcal{L}=\alpha \frac{\|\theta^{\rm t}-\hat{\theta}^{\rm t}\|_2^2}{\|\theta^{\rm t}\|_2^2+\epsilon}+\beta{\|\theta^{\rm r}-\hat{\theta}^{\rm r}\|}_2^2
\end{equation}
where $\|\cdot\|_2:=\sqrt{(\sum\cdot)^2}$ is Euclidean norm, the both weights of relative translation error $\alpha$ and rotation error $\beta$ are set as $\alpha=\beta=0.5$, and $\epsilon=0.01$ to avoid singularity.
Momentum Stochastic Gradient Descend 
was applied with the learning rate 0.0001 and the learning momentum 0.9.

We split our CT data set into two folders (A and B) each containing pairs of contrast and non contrast-enhanced CTs from the same mouse, and two-fold cross validation was performed on different mouse. Furthermore, we performed four validation strategies: (S1) training on all data from folder A, and testing on B; (S2) training on all data from B, testing on A; (S3) training on non contrast-enhanced data from A, and testing on all data from B; (S4) training on non contrast-enhanced data from B, and testing on all data from A. Training on non contrast-enhanced data was performed to check whether our network can be used for alignment of follow-up contrast-enhanced data, when only the baseline data are available for training, thus modeling real scenario of data acquisition. For the validation strategy, known transformation (as described in Sec~\ref{sec:data}) was applied to volume from the folder to create a pair of CTs in synthetic test, and in real test, each pair of corresponding contrasted and non-contrasted CT was registered. The synthetic test includes rotation test with fixed translation $(0.4~0.4~0.4)$mm and 11 angles rotation uniformly ranges from $-\pi$ to $\pi$ around axis $(\frac{1}{\sqrt{3}}~\frac{1}{\sqrt{3}}~\frac{1}{\sqrt{3}})$ as well as translation test with fixed $\frac{\pi}{2}$ rotation around axis $(\frac{1}{\sqrt{3}}~\frac{1}{\sqrt{3}}~\frac{1}{\sqrt{3}})$ and 11 translation uniformly ranges from $-\frac{\sqrt{3}}{2}$mm to $\frac{\sqrt{3}}{2}$mm along the axis $(\frac{1}{\sqrt{3}}~\frac{1}{\sqrt{3}}~\frac{1}{\sqrt{3}})$. 

\subsection{Comparison and evaluation}
\label{sec:comp}
D-net was compared with other relevant image registration approaches:
\\$\bullet$ {SITK}: Simple ITK with metric Joint Histogram Mutual Information and optimizer Regular Step Gradient Descent with gradient tolerance 0.0001, max iter number 10k and learning rate 1.
\\$\bullet$ {ME (Mixed Encoder)}: The "Global-net" \cite{hu2018label} concatenating the two input volumes together and feeding into one mixed branch. All the architecture settings are set with default values, with $\textit{\textbf{d}}=\textit{\textbf{h}}=\textit{\textbf{w}}=(64,32,16,8,4)$ and $\textit{\textbf{c}}=(1,16,32,64,128)$.
\\$\bullet$ {SE (Siamese Encoder)}: An architecture employing two branches of 6 Res-down blocks for SE and two fully connected layers for regression, a similar structure was used in \cite{de2019deep} but without residual structure and fewer down-sampling blocks.
\\$\bullet$ {SED (Siamese Encoder Decoder)}: A proposed architecture inserting 4 Res-up blocks into SE between the SE and regression parts, with skip connection from the 4 latter Res-down block of SE.
\\$\bullet$ {SNL-SED (Self Non-local Linked - Siamese Encoder Decoder)}: A proposed SED architecture with the 4 latter Res-down blocks self non-locally linked by the Embedded Gaussian similarity based - non-local block \cite{wang2018non} in each branch.

SITK, ME and SE are previously published methods; while SED and SNL-SED are transitional forms towards D-Net and their impact is separately validated; SE, SED SNL-SED and D-net are validated with $\textit{\textbf{d}}=\textit{\textbf{h}}=\textit{\textbf{w}}=(64,32,16,8,4,2,1)$ and $\textit{\textbf{c}}=(1,16,32,64,64,64,64)$.

\subsubsection{Criteria}
The Euclidean distance of Translation Error (TE) between the predicted and expected translation, $TE:={\|\theta^{\rm t}-\hat{\theta}^{\rm t}\|}_2$, and the Rotation Error (RE) between the predicted and expected rotation, $RE=\arccos(\frac{{\rm tr}(\textit{\textbf{R}}^\top O(\hat{\theta}^{\rm r}))-1}{2})$, are calculated for synthetic tests. Since the ground truth for real examples are unknown, the Dice Similarity Coefficient (DSC) between the cortical bone segmented from contrasted and non-contrasted tibial CT is calculated for both synthetic and real tests. 

\begin{figure}[ht]
\includegraphics[width=\textwidth]{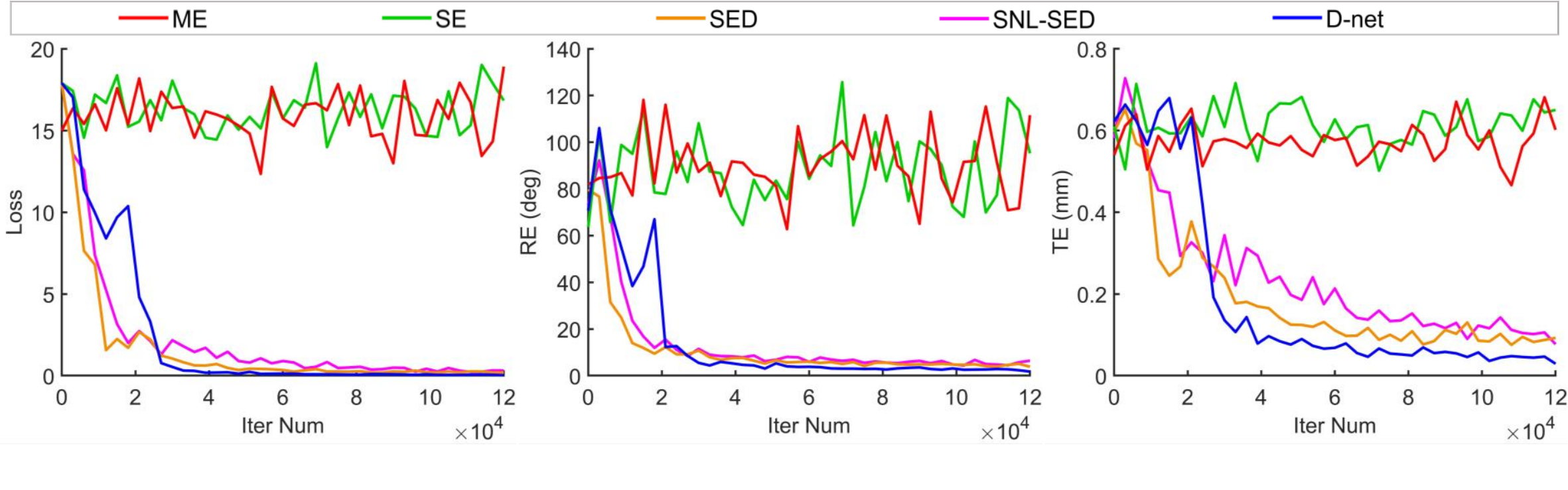}
\caption{The training curves exemplified with (S1) left: Loss values, middle: Rotation errors (RE) and right: Translation Errors (TE) shows SED, SNL-SED and D-net are trainable across range of translations and rotations.} \label{fig:curve}
\end{figure}

\section{Results}
\label{sec:result}

\begin{figure}[ht]
\includegraphics[width=\textwidth]{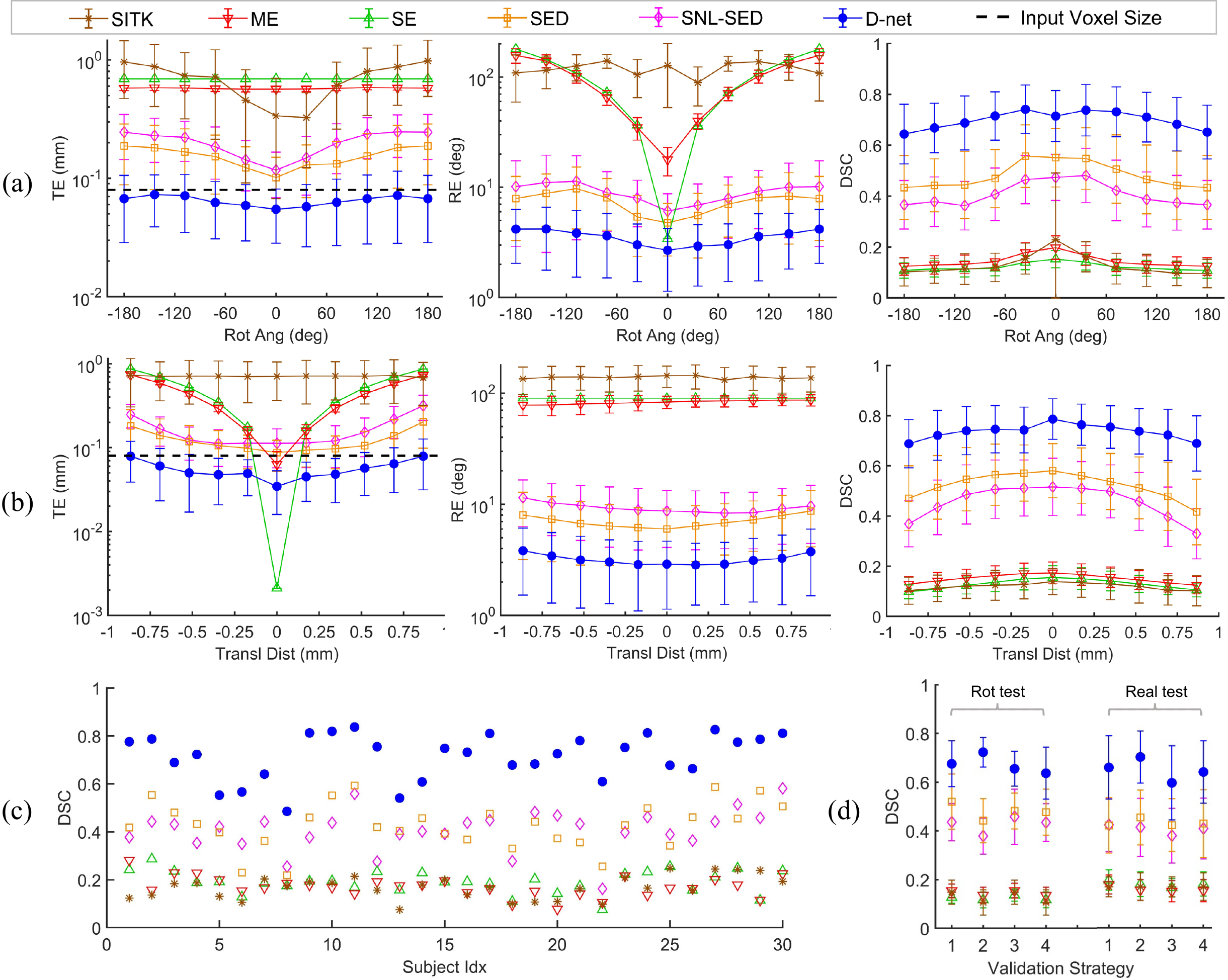}
\caption{D-net outperforms all other methods exemplified by (S1) in (a) rotation (rot) test, (b) translation test, and (c) real test exemplified by 30 subjects, with Translation Errors (TE), Rotation Errors (RE) and DSC, avg$\pm$std; (d) DSC, avg$\pm$std in rot test and real test with validation strategies (S1)$\--$(S4).} \label{fig:results}
\end{figure}

\begin{figure}[ht]
\includegraphics[width=\textwidth]{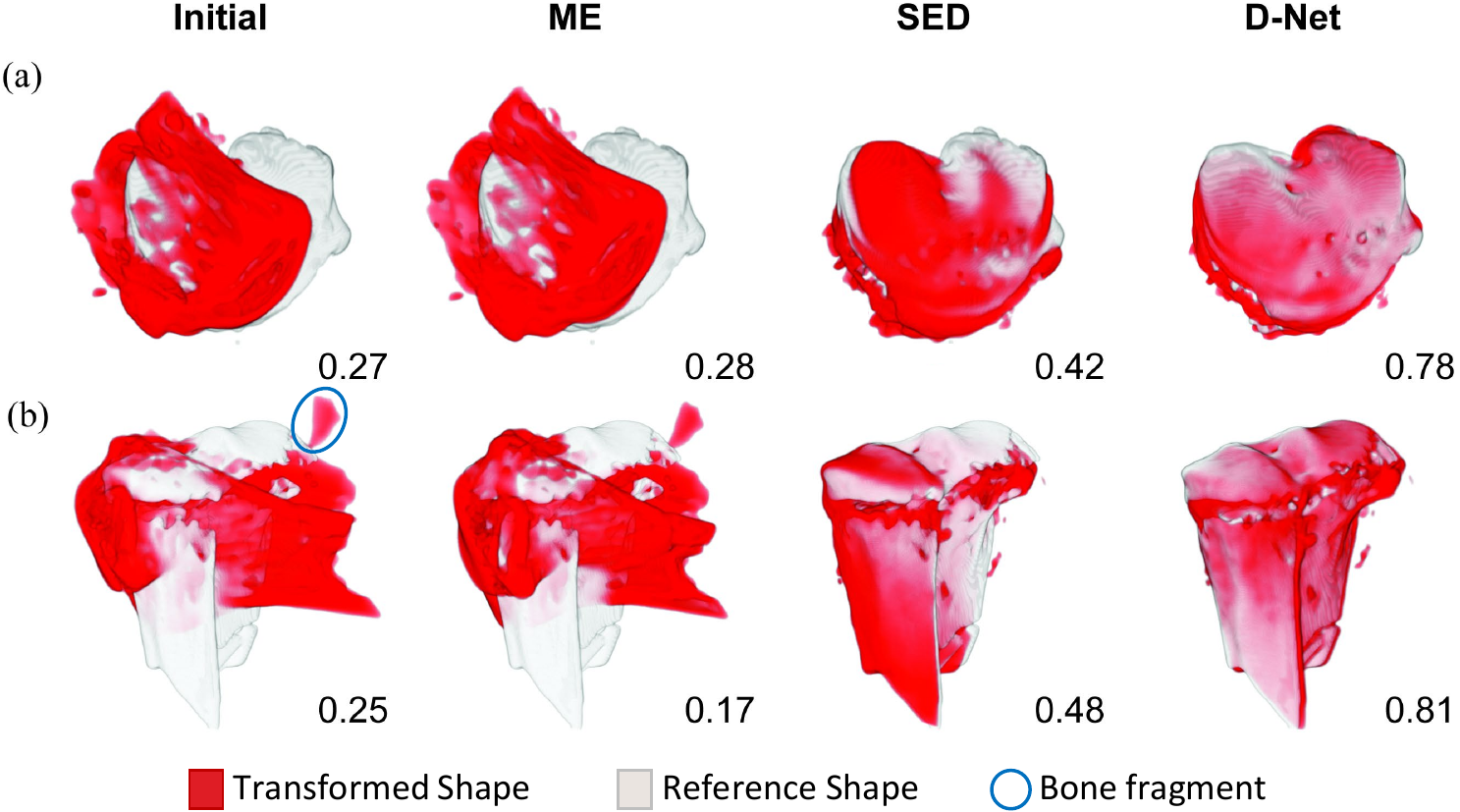}
\caption{Segmentation surfaces for the two exemplar volumes used in real experiment. D-net achieve the most plausible registration (overlapping red and white surfaces) with highest DSC shown at the right bottom corner (images are shown at original resolution).}
\label{fig:examples}
\end{figure}

All networks were trained for 120k iterations. In all training strategies used, ME and SE failed to converge, whereas SED, SNL-SED and D-net were trainable (exemplified in Fig.~\ref{fig:curve}).

The results of rotation test and translation test with validation strategy (S1\&S2) are shown in Fig.~\ref{fig:results}(a) and Fig.~\ref{fig:results}(b), where only D-net achieves the sub-voxel average TE in the rotation and translation tests. The performance of ME and especially SE is sensitive to the initial translation and rotation as shown in Fig.~\ref{fig:results}(a)-middle and Fig.~\ref{fig:results}(b)-left because they intend to predict a small range transformations for any input volumes. DSCs for 30 subjects in real test with strategy (S1\&S2) are shown in Fig.~\ref{fig:results}(c), where the SED increases DSC by 0.1$\--$0.5 from ME and D-net further raises DSC by 0.1$\--$0.4 from SED. The average DSC for strategy (S1)$\--$(S4) in rotation test and real test are plotted in Fig.~\ref{fig:results}(d). It shows average DSC of D-net is higher than all others in both rotation and real test but is slightly lower in real test compared with rotation test. 

The tibial bone shapes of two registration examples for ME, SED, and D-net from real test are shown in Fig.~\ref{fig:examples} with the same subjects previously shown in Fig.~\ref{fig:ct}. The figure illustrates bone fragments and segmentation difference caused by the varying intensity influenced by contrast, decreasing the DSC values and making registration of preclinical tibia data particularly challenging. Visual results in Fig.~\ref{fig:examples} confirms that D-Net performs robustly and this is further supported by the quantitative results shown in Tab.~\ref{tab1}, where TE, RE, and DSC for all methods are presented. 

Comparing with others, D-net achieves the lowest TE and RE and highest DSC with consistent performance across range of rotations. Using two-way Analysis of Variance (ANOVA) in rotation and translation tests and one-way ANOVA in real test, D-net significantly outperforms all other approaches with $p<10^{-4}$ on TE, RE and DSC in rotation and translation tests and on DSC in real test by strategy (S1\&S2) and (S3\&S4); SED significantly outperforms SE, ME and SITK with $p<10^{-4}$ on TE, RE and DSC in rotation and translation test and on DSC in real test by all the strategies.

\begin{table}[h!]
\caption{Average values of Translation Error (TE/$\mu$m), Rotation Error (RE/$^\circ$) and Dice Similarity Coefficient (DSC/\%) for different methods in Translation (Transl), Rotation (Rot) and Real test, with strategies of training on both contrasted and non-contrasted data (S1\&S2) and just non-contrasted data (S3\&S4)}
\label{tab1}
\begin{tabular}{ccC{0.92cm}cccccccccc}
\hline
&&SITK&\multicolumn{2}{p{1.74cm}<{\centering}}{ ME}&\multicolumn{2}{p{1.74cm}<{\centering}}{ SE}&\multicolumn{2}{p{1.74cm}<{\centering}}{SED}&\multicolumn{2}{p{1.74cm}<{\centering}}{SNL-SED}&\multicolumn{2}{p{1.74cm}<{\centering}}{D-net} \\

\multicolumn{2}{c}{Variable No.}&-&\multicolumn{2}{p{1.74cm}<{\centering}}{0.7M}&\multicolumn{2}{p{1.74cm}<{\centering}}{9.2M}&\multicolumn{2}{p{1.74cm}<{\centering}}{5.2M}&\multicolumn{2}{p{1.74cm}<{\centering}}{4.9M}&\multicolumn{2}{p{1.74cm}<{\centering}}{4.9M}\\

\multicolumn{2}{c}{Strategy(S)}&-&1\&2&3\&4&1\&2&3\&4&1\&2&3\&4&1\&2&3\&4&1\&2&3\&4\\
\hline
\multirow{3}{*}{Transl} &TE&710.5  &406.4&395.1&472.6 &472.5&124.3 &132.7&163.4&146.2&\textbf{55.8}&\textbf{77.7}\\
 &RE&135.9  &82.7&83.4 &89.8&89.2 &7.1&8.0&9.3&9.7&\textbf{3.2}&\textbf{4.2}\\
&DSC&12.98  &15.04& 14.94&12.86&13.42 &52.28&51.08&45.61&47.69&\textbf{73.57}&\textbf{66.44}\\\hline
\multirow{3}{*}{Rot} &TE  &696.2 &575.4&569.4 &692.4&692.4&154.4&144.1&201.5&175.7&\textbf{65.3}&\textbf{81.0}\\
 &RE  &119.6 &92.6 &95.1 &98.5&98.0&7.4&8.1&9.1&9.5&\textbf{3.6}&\textbf{4.5}\\
&DSC  &12.75 &14.54 &14.54&12.22 &12.75&48.28&47.97&40.88&44.69&\textbf{69.83}&\textbf{64.64}\\\hline
\multirow{1}{*}{Real} &DSC &16.91 &16.75&15.16&18.86 &17.23&43.66 &42.52&41.85&39.26&\textbf{67.92}&\textbf{61.68}\\
\hline
\end{tabular}
\end{table}

\section{Discussion and Conclusion}
\label{sec:discussion_conclusion}
In this paper, we proposed a new network, D-net, and a new structure, Mutual Non-local Link (MNL), for estimation of transformation between CT volumes.

The experimental results shows D-net outperforms other methods and achieves state-of-the-art performance for rigid registration of preclinical mouse CT scans with and without contrast. While ME and SE did not converge during training using full range of rotations, we were able to train them using smaller range of rotations (30$^\circ$), similarly as in \cite{chee2018airnet,haskins2019learning}. This further shows superiority of D-net at consistently estimating full range of rotations. 

The average DSCs of D-net in real test are slightly lower than in synthetic tests potentially due to the difference in segmentation for contrast-enhanced volumes. 
However, D-net is still able to extract the common features of tibial bone and align two volumes plausibly, showing usefulness for shape extraction of cartilage from contrast-enhaced CT of tibiae.

For the rotation representation, the widely used quaternion, Euler angles and Lee algebra were not applied due to the discontinuity of 3D rotation represented in the real Euclidean space with dimension lower than 5D \cite{zhou2019continuity}.

In future work, a pipeline for cartilage shape extraction will be further validated for morphological analysis and in application for diagnosis and staging of osteoarthritis, and D-net will be generalized to other modalities to explore inter-subject and inter-modality registration.


\section{Acknowledgement}
This work was supported by a Kennedy Trust for Rheumatology Research Studentship, the Centre for OA Pathogenesis Versus Arthritits (Versus Arthritis grant 21621). The authors acknowledge Patricia das Neves Borges as the researcher who collected the preclinical CT dataset, as part of the National Centre for Replacement, Refinement and Reduction of Animals in Research (NC3R grant NC/M000141/1).B. W. Papież acknowledges Rutherford Fund at Health Data Research UK

%
%
%
%
%
\bibliographystyle{splncs04}
\bibliography{reference}
%




\end{document}